\documentclass[12pt]{article}
\usepackage{epsfig}
\newcommand{\nlabel}[1]{\label{#1}}
\newcommand{\be}{\begin{equation}}
\newcommand{\ee}{\end{equation}}
\newcommand{\bea}{\begin{eqnarray}}
\newcommand{\eea}{\end{eqnarray}}
\newcommand{\ba}{\begin{array}}
\newcommand{\ea}{\end{array}}
\begin{document}
\title{\bf Out of Equilibrium Non-perturbative Quantum Field Dynamics in
Homogeneous External Fields \\}
\author{{\bf F. J. Cao,  H. J. de Vega} \\ \\
LPTHE, Universit\'e Pierre et Marie Curie (Paris VI) et \\ Denis Diderot
(Paris VII), Tour 16, 1er. \'etage, 4, Place Jussieu, \\ 75252 Paris,
Cedex 05, France}
\date{\today}
\maketitle
\begin{abstract}
The quantum dynamics of the symmetry broken $ \lambda\,
({\vec{\Phi}^2})^2 $ scalar
field theory in the presence of an homogeneous external field is
investigated in the large $ N $ limit. We choose as initial state
the ground state for a constant external field $ \vec{\cal J} $. The
sign of the external field is suddenly flipped from $ \vec{\cal J} $
to  $ - \vec{\cal J} $ at a given time and the
subsequent quantum dynamics calculated. Spinodal instabilities and
parametric resonances produce large quantum fluctuations in the field
components transverse to the external field. This allows the order
parameter to turn around the maximum of the potential for intermediate
times. Subsequently, the order parameter starts to oscillate near the
global minimum for external field $ - \vec{\cal J} $, 
entering a novel quasi-periodic regime.
\end{abstract}
\tableofcontents

\section{Introduction}

The understanding of the physics at large energy density situations
is fundamental to achieve a proper picture both
for ultrarelativistic heavy ion collisions \cite{refexp}
and for the early universe \cite{higheuni, noscos, tsuinf}.
These physical phenomena cannot be treated with the usual quantum field
perturbation methods and call for self-consistent nonperturbative
methods as the large $N$ approach and Hartree approximations
\cite{higheuni, noscos, tsuinf, renorm, j0eta, tsunos}.

Here, we study effects of external fields in non-perturbative
quantum field dynamics using the large $ N $ limit method.

The effective potential of the $O(N)$ $ \lambda\,\vec\Phi^4$ theory with
broken
symmetry has two degenerate minima for zero external field. In the
presence of a small  uniform external field $ \vec{\cal J} $ the
potential
becomes tilted and the degeneracy between the minima lifted. The
absolute minimum becomes the vacuum while the other local minimum
becomes a  metastable vacuum. See Fig. \ref{fclaspotljcini}.

We take as initial state the ground  state. That is, the field
expectation value is initially at the absolute minimum
and the field modes are in the vacuum.
Then, at a given time we change the sign of
the external field: $ \vec{\cal J} \rightarrow -\vec{\cal J}
$. Flipping the sign of the external field exchanges the local minima.
That
is, the absolute minimum becomes the false vacuum and viceversa.
As a consequence, the system is now near the false vacuum.
Spinodal resonances develop and the fluctuations
transverse to the direction of $ \vec{\cal J} $ grow exponentially
till they are shut-off by the non-linear back-reaction. The
expectation value of the field  tends  to reach the absolute
minimum which is on the other side of a potential barrier for values
$ |\vec{\cal J}| $  smaller than
$ {2 \over \sqrt{27}} \; (|m|^3 \, \sqrt{2\over\lambda}) $.
See Fig. \ref{fclaspotljcini}.

There are two regimes for the dynamics separated by a critical value of the
external field $ J_c =  \sqrt{ 2 \; {\left( 13^2 + 15 \sqrt{5} \right) \over
19^3}}  \; (|m|^3 \, \sqrt{2\over\lambda}) $:

For $ |\vec{\cal J}| > J_c $ the system
quickly reaches a quasiperiodic regime where the field expectation
value oscillates near the true quantum vacuum.
See Figs. \ref{fclaspotgjc}, \ref{fetagjc}, \ref{fgsgjc} and
\ref{fm2gjc}.

For $ |\vec{\cal J}| < J_c $ the field cannot reach the absolute
minimum moving in the longitudinal direction  except by tunnel effect
which is suppressed in
the large $N$ limit. Instead, the field turns around the potential
maximum developing large transverse quantum fluctuations
$ {\cal O}(m^2/\lambda) $.
It takes a time $ t_s \sim {1 \over \sqrt{j} \, m} \log{1 \over
\lambda} $
(spinodal time) for the quantum fluctuations to become of the order $
{\cal O}(m^2/\lambda) $. By this time the field expectation value makes
a spectacular jump and starts to oscillate  near the true vacuum. See
Figs. \ref{fclaspotljc}, \ref{fetaljc} and \ref{fgsljc}.

For later times the evolution in both cases becomes quasiperiodic
showing a clear
separation between fast and slow variables. The fast time dependence
of the field expectation value is explicitly found in terms of
elliptic Jacobi functions. Comparison with the full numerical solution
shows that their parameters exhibit slow time dependence.
The smaller is $  |\vec{\cal J}| $ the slower are the slow
variables.
We also see that the mass squared is of order $  |\vec{\cal J}| $.
For $ |\vec{\cal J}| = 0 $ we recover the known zero
external field results \cite{renorm,j0eta}.

We consider small values of the coupling constant $ \lambda \ll 1 $
in order to separate the different time scales of the evolution.

\section{Classical field dynamics}  \nlabel{sclasevol}

We will first propose and solve the problem in classical field theory,
in order to present results that will be used later for comparison.
We consider the classical $ O(N) $-invariant scalar field model with quartic
self-interaction \cite{tsunos} in the presence of an homogeneous
external
field $ {\vec{\cal J}} $. We study in this model the evolution of the
ground state after the change $ \vec{\cal J} \rightarrow -\vec{\cal J} $
in the external field.

\subsection{Classical evolution equation and initial conditions}
\nlabel{ssclasevol}

The action and Lagrangian density are given by,
\begin{eqnarray}
S  &=&  \int d^4x\; {\cal L}\; \; ,\nlabel{clasaction} \cr \cr
{\cal L}  &=&   \frac{1}{2}\left[\partial_{\mu}{\vec{\Phi}}(x)\right]^2
-\frac12\;m^2\;{\vec{\Phi}^2}
- \frac{\lambda}{8N}\left({\vec{\Phi}^2}\right)^2 + \vec{\cal
J}\cdot\vec{\Phi} \;.
\nlabel{claslagrangian}
\end{eqnarray}
We restrict ourselves to a translationally invariant situation.
In this case the field $ {\vec \Phi} $ and the external field
$ \vec{\cal J} $ are independent of the spatial coordinates $ \vec{x} $
and only depend on time. The evolution equation is then
\be
\ddot{\vec\Phi}(t) + \left( m^2 + \frac{\lambda}{2N}\;\vec{\Phi}^2(t)
\right)
\vec{\Phi}(t) = {\vec{\cal J}}(t) \nlabel{clasevol}
\ee
and the classical potential
\be
V({\vec\Phi}) = \frac12\;m^2\;{\vec{\Phi}^2} +
\frac{\lambda}{8N}\left({\vec{\Phi}^2}\right)^2 - {\vec{\cal
J}}\cdot\vec{\Phi} \;.  \nlabel{claspot}
\ee
We consider the broken symmetry case ($ m^2 < 0 $).

Choosing the first axis of the internal space in the direction of the
external field $ {\vec{\cal J}} $, their components are
$ {\vec{\cal J}} = (\sqrt{N}J,\, 0,\,\, \ldots\,,\, 0) $.
Consequently, the $ \vec{\Phi} $ field can be expressed as
$ \vec{\Phi} = (\sqrt{N}\phi,\, \vec\phi_\pi) $.

Thus, the potential becomes:
\bea \label{potef}
V({\vec\Phi}) &=& N\; V_J(\phi) + V_\pi(\vec\phi_\pi) \cr \cr
V_J(\phi) &=& \frac12\;m^2\;\phi^2 + \frac{\lambda}{8}\;\phi^4 - J\;
\phi \\
V_\pi(\phi,\vec\phi_\pi) &=& \frac12\;m^2\;{\vec{\phi_\pi}^2}
  + \frac{\lambda}{8N}\left[2\;\phi^2\;\vec{\phi_\pi}^2
  + \left({\vec{\phi_\pi}^2}\right)^2 \right]  \;. \nonumber
\eea

We choose the \emph{ground state as initial state}, therefore it has the
form
(this can be shown from the previous expressions)
\bea
&&\phi(t_i) = \phi_0 \quad ; \quad \dot\phi(t_i) = 0  \cr \cr
&&\vec{\phi_\pi}(t_i) = 0 \quad ; \quad \dot{\vec{\phi_\pi}}(t_i) = 0
\label{condit}
\eea
where $ \phi_0 $ is the global minimum of $ V_J(\phi) $.

The evolution equations for $ \vec{\phi_\pi} $ are
\be
\ddot{\vec{\phi_\pi}}(t) +
\left[ m^2 + \frac{\lambda}{2N}
\left(\phi^2(t)+\vec{\phi_\pi}^2(t)\right)\right] \vec{\phi_\pi}(t) = 0
\;,
\ee
and with the previous initial conditions, \emph{independently} of the
value
of $ J $, the solution is simply
\be
\vec{\phi_\pi}(t) = 0 \;.
\ee
Therefore, the \emph{classical} problem reduces to the following
\emph{one component} field problem.
The evolution equation for $ \phi $ reduces to
\be\label{eqevolfi}
\ddot\phi(t) + \left[ m^2 + \frac{\lambda}{2}\;\phi^2(t) \right] \phi(t)
= +J \; \quad \mbox{for }\; t<0 \; .
\ee
As it is obvious, the ground state $ \phi_0 $ is a constant solution of
this
equation. We see from Eq. (\ref{potef}) that $ \phi_0 $, the global
minimum of $
V_J(\phi_0) $ fulfills Eq. (\ref{eqevolfi}).

\bigskip

We introduce the dimensionless variables
\be \label{sindim}
\tau \equiv |m|\,t \quad ; \quad \eta \equiv \sqrt{\lambda \over 2} {
\phi\over |m|} \quad , \quad
j \equiv \sqrt{\lambda \over 2} { J \over |m|^3}\quad , \quad
V_{\tau \leq 0} \equiv \frac{\lambda}{2\,m^4}\,V_J \quad , \quad
\vec{ j} \equiv \sqrt{\lambda \over 2 \, N} { 1 \over |m|^3} \;
\vec{\cal J}
\quad .
\ee
Therefore,
$$
V_{\tau \leq 0}(\eta) = -\frac12\;\eta^2 + \frac{1}{4}\eta^4 - j \eta \;
,
$$
since we choose $ m^2 < 0 $.

Thus, the classical vacua are the solutions of the cubic equation $
V'_{\tau \leq 0}(\eta) = 0 $,
\be \label{cubica}
\eta^3 - \eta - j = 0
\ee
For small external fields
$ j < {2 \over \sqrt{27}} = 0.3849002\ldots $ Eq. (\ref{cubica}) has
three real
roots. The ground state (the global minimum) of the potential
$ V_{\tau \leq 0} $ is at
\be \label{eta1}
\eta = \eta_0 \equiv {2 \over \sqrt3 }\cos\left[\frac13
\mbox{arccos}\left(
{ \sqrt{27} \, j \over 2}\right) \right] = 1+\frac12 j -\frac38 j^2
+\frac12j^3 -\frac{105}{128}j^4 + O(j^5) \;,
\ee
there is a local minimum (false vacuum) at
$$
\eta_{fv} \equiv {2 \over \sqrt3 }\cos\left[\frac13 \mbox{arccos}\left(
{ \sqrt{27} \, j \over 2}\right) + {2 \pi \over 3} \right] = -1 +
\frac12 j
+\frac38 j^2 + \frac12j^3
  +\frac{105}{128}j^4 + O(j^5) \;,
$$
and a local maximum at
\be \label{max}
\eta_M \equiv {2 \over \sqrt3 }\cos\left[\frac13 \mbox{arccos}\left(
{ \sqrt{27} \, j \over 2}\right) + {4 \pi \over 3} \right] =-j -j^3 +
O(j^5)
\ee

For larger external fields $ j > {2 \over \sqrt{27}} $ there is only
one real minimum, (the absolute minimum) at
\bea
\eta = \eta_0 &\equiv& \left( {j \over 2} \right)^{1/3}\left[ \left(1 +
\sqrt{
  1 - {4  \over 27 \, j^2}}\right)^{1/3} + \left(1 - \sqrt{
  1 - {4  \over 27 \, j^2}}\right)^{1/3}\right]  \cr\cr
&=& j^{1/3} + { 1 \over 3 \,
  j^{1/3}} - { 1 \over 81 \, j^{5/3}} + O(j^{-7/3}) \; . \nonumber
\eea
We shall concentrate in the more interesting case $ j < {2 \over
\sqrt{27}} $.

The field stays in the ground state $ \eta_0 $ with $ j > 0 $ from the
(negative) initial time $ \tau_i $ till  $ \tau = 0 $. At $ \tau = 0 $
we suddenly flip the sign of the external field $ j $.

That is, the time dependence of the external field is
\be
\vec{j} = \left\{ \ba{ll}
(j,\, 0,\,\, \ldots\,,\, 0)  & \mbox{ for } \tau \le 0 \\
(-j,\, 0,\,\,\ldots\,,\, 0) & \mbox{ for } \tau > 0 \\
\ea \right. \nlabel{clasJt}
\ee
This change of sign of $ j $ introduces some amount of energy in the
system.

As previously stated this problem reduces to a one component field
problem.
Therefore, the potential is given in dimensionless variables by (see
Fig. \ref{fclaspotgjc},
\ref{fclaspotljc})
\be
V(\eta) = \left\{
\ba{llll}
V_{\tau \leq 0}(\eta)  & \equiv & -\frac12\;\eta^2 + \frac{1}{4}\eta^4 -
j \eta
& \mbox{ for } \tau \le 0 \\
V_{\tau>0}(\eta) & \equiv & -\frac12\;\eta^2 + \frac{1}{4}\eta^4 + j
\eta
& \mbox{ for } \tau > 0
\ea \right.   \nlabel{Vj}
\ee
The equation of motion for $ \tau > 0 $ is
\be
\ddot\eta + \left( -1 + \eta^2 \right)\eta = -j \; ,
\ee
where $ -1  + \eta^2 $ plays the role of an effective mass.
The initial state at $ \tau  = 0 $ is the ground state for $
\tau \leq 0 $, and it has the form
\be
\eta(0) = \eta_0
\quad ; \quad \dot\eta(0) = 0 \;.
\ee

\subsection{The classical dynamics}

We have shown above that the initial state (the ground state for an
external
field $ j $) is a stationary state of the evolution equations for
$ \tau \le 0 $. The change $ j \rightarrow -j $ at  $ \tau= 0 $ breaks
this
stationarity and the system now finds itself near a metastable state
(see Fig. \ref{fclaspotgjc}, \ref{fclaspotljc}).
We consider $ j < {2 \over \sqrt{27}} $ when there is a potential
barrier (see Fig. \ref{fclaspotljcini}).
This change in the sign of the external field $ j $, increases the
energy
density of the system by an amount $ V_{\tau>0}(\eta_0) - V_{\tau \leq
0}(\eta_0) = 2\,j\,\eta_0 $. However, even if the field is near
the metastable minimum, it could have now enough energy to jump above
the
barrier and reach the global minimum of the potential.
As the height of the barrier is $ V_{\tau>0}(\eta_{M'}) $ the condition
to jump  above the barrier is
\be
V_{\tau>0}(\eta_0) > V_{\tau>0}(\eta_{M'}) \; ,
\ee
where $ \eta_{M'} $ is the local maximum of the potential $
V_{\tau>0}(\eta) $.

Solving this equation one obtains that the field jumps  above the
barrier provided $ |j| > j_c $ with
\be
j_c = \sqrt{ 2 \; {\left( 13^2 + 15 \sqrt{5} \right) \over 19^3}} =
0.243019\ldots \;.
\ee
[Notice that $ j_c <  {2 \over \sqrt{27}} $ and therefore there is
indeed a barrier for these values $ j \leq j_c $.]

This critical value $ j_c $ of the external field separates the
dynamics in two  regimes
\begin{itemize}
\item $ |j| > j_c $. The field jumps  above the barrier and reaches the
region
of the global minimum. After this the field continues to jump  above
the barrier back and forth with periodic oscillations. See Fig.
\ref{fclaspotgjc} and dashed line in Fig. \ref{fetagjc}.

\item $ |j| < j_c $. The system does \emph{not} jump  above the barrier,
and oscillates periodically between the turning points of the motion $
\eta_0 $ and $\eta_t$ around the metastable minimum.
$ \eta_0 $ is given by Eq. (\ref{eta1}) and $\eta_t$ is the solution of
the
equation
$$
V_{\tau>0}(\eta_t) = V_{\tau>0}(\eta_0)
$$
we find,
\be
\eta_t = 1 - \frac32 j - \frac{11}{8}j^2 - \frac72 j^3
   - \frac{1049}{128} j^4 + O(j^5) \nlabel{clasturnp}
\ee
Since both $ \eta_t $ and $ \eta_0 $ are positive, $ \eta $ is always
positive.
See Fig. \ref{fclaspotljc} and  dashed  line in Fig. \ref{fetaljc}.

\end{itemize}

\section{Quantum field dynamics}

In this section we derive the quantum equation of motion and compute
the quantum field dynamics using the large $ N $ limit.
As in the previous section, we consider the $ O(N) $-invariant scalar
field
model with quartic self-interaction \cite{tsunos} in the presence of an
homogeneous external field $ {\vec{\cal J}} $. We study in this model
the evolution of the ground state after the change
$ \vec{\cal J} \rightarrow -\vec{\cal J} $ in the external field.

\subsection{Quantum equations of motion and initial conditions}
\nlabel{squantumeq}

The action and Lagrangian density are,
\begin{eqnarray}
S  &=&  \int d^4x\; {\cal L}\; \; ,\nlabel{action} \cr \cr
{\cal L}  &=&   \frac{1}{2}\left[\partial_{\mu}{\vec{\Phi}}(x)\right]^2
-\frac12\;m^2\;{\vec{\Phi}^2}
- \frac{\lambda}{8N}\left({\vec{\Phi}^2}\right)^2
+ \vec{\cal J}\cdot\vec{\Phi}
\nlabel{lagrangian}
\end{eqnarray}
where $ \vec{\Phi} $ is now a quantum operator.

We restrict ourselves to a translationally invariant situation.
In this case  the order parameter $ \langle {\vec \Phi}(\vec{x}, t)
\rangle $
and the external field $ \vec{\cal J}(\vec{x}, t) $ are independent of
the
spatial coordinates $ \vec{x} $ and only depend on time. Moreover, we
choose the direction of $ \vec{\cal J}(t) $ independent of time. Then,
we can choose this direction as the first axis in the $N$-dimensional
internal space.
\begin{eqnarray}
\vec{\cal J} &=& \left\{
  \ba{ll}
    (J,\, 0,\,\, \ldots\,,\, 0)  & \mbox{ for } t \le 0 \\
    (-J,\, 0,\,\,\ldots\,,\, 0) & \mbox{ for } t > 0 \\
  \ea \right. \nlabel{Jt} \; .
\end{eqnarray}
The field expectation value in the problem we are studying
is parallel to the external field:
\be
\vec{\Phi}(x) = (\sigma(x), \vec{\pi}(x) )
  = ( \sqrt{N} \phi(t)+\chi(x), \vec{\pi}(x) ) \nlabel{split}
\ee
with $ \phi(t) = \langle \sigma(x) \rangle $; thus,
$ \langle \chi(x) \rangle = 0 $. While in the $ N-1 $ directions
transversal
to the expectation value we have $ \langle {\vec \pi}(\vec{x},t) \rangle
=0 $.

The derivation of the equations of motion in the large $ N $ limit in
this
case, is analogous to the case without external field, that has been
explained in detail in \cite{tsunos}.
Thus, we will only present here the main concepts and equations.

We have one direction parallel to the expectation value, and
$ N-1 $ transversal directions. The transverse fluctuations dominate
in the large $ N $ limit while the longitudinal fluctuations only
contribute to the $ 1/N $ corrections to the equations of motion.

The evolution equation for the expectation value is for $ t > 0 $
\be
{\ddot \phi}(t) + \left\{ m^2 + \frac{\lambda}{2}\left[ \phi^2(t)
  + \frac{\langle \vec{\pi}^{\,2}(x) \rangle}{N} \right] \right\}\phi(t)
= -J
\ee
where the dot denotes the time derivative. We see in the previous
equation
that
$$
{\cal M}_d^2(t) \equiv m^2 + \frac{\lambda}{2}\left[\phi^2(t)
+ \frac{\langle \vec{\pi}^{\,2}(x) \rangle}{N} \right]
$$
plays the role of an effective mass.
The last term in $ {\cal M}_d^2(t) $ has a quantum origin and it is
absent
in the classical effective mass. It can be interpreted as an in-medium
effect due to the presence of $ \vec \pi $ particles.

In the Heisenberg picture we can write
\be
\vec{\pi}(\vec x,t) = \int \frac{d^3{\vec k}}{\sqrt{2} (2\pi)^3}\;
\left[ \vec{a}_k\; \varphi_k(t) \;
e^{i\vec k \cdot \vec x}+\vec{a}^{\dagger}_k \;\varphi^*_k(t)\;
e^{-i\vec k \cdot \vec x} \right]
\nlabel{fieldexpansion}
\ee
where $ \vec{a}_k \; , \; \vec{a}^{\dagger}_k $ are the
time independent annihilation and creation operators with the usual
canonical  commutation relations.
Thus, the $ \varphi_k(t) $ are the mode functions of the field and we
can express $ \langle\vec{\pi}^{\,2}(x)\rangle $ as
\bea
\frac{\langle\vec{\pi}^{\,2}(x)\rangle}{N}
  &=& \frac{1}{2} \int{ \frac{d^3 k}{(2\pi)^3}\; \left[|\varphi_k(t)|^2
  - {\cal S}_d \right] } \; . \nlabel{fluctmode} \cr \cr
{\cal S}_d &=& \frac{1}{k} - \frac{\theta(k-\kappa)}{2k^3}\; {\cal
M}_d^2(t)
\eea
where $ \kappa $ is an arbitrary renormalization scale, and we will
choose
$ \kappa = |m_R| $ for simplicity. (For details on the
renormalization procedure, that leads to the substraction $ {\cal S}_d $
see Ref. \cite{renorm}.)

In the previous expressions we have written $ k $, instead of $ \vec k
$, since we choose  spherically symmetric initial conditions.

The mode functions have the following evolution equations
\be
\ddot{\varphi}_k(t)+\omega_k^2(t)\;\varphi_k(t)=0\quad ,  \quad
\omega_k^2(t) \equiv k^2 + {\cal M}_d^2(t)  \; .
\ee
We consider values of the field
$ |J| < {2 \over \sqrt{27}} \; (|m|^3 \, \sqrt{2\over\lambda}) $
where the potential $ V_J $ has two local minima.
As initial state we take the ground state for $ t \leq 0 $
\bea
\phi(0) &=& \phi_0 \quad\quad\quad ; \quad\quad\quad \dot \phi(0) = 0
\cr \cr
\varphi_k(0) &=& \frac{1}{\sqrt{\omega_k(0)}}  \quad ; \quad
  \dot \varphi_k(0) = -i\; \sqrt{\omega_k(0)} \;.
\eea
where $ \phi_0 $ is the global minimum of $ V_J(\phi) $ [defined in
Eqs. (\ref{sindim}) and (\ref{eta1})] and
$ \omega_k(0) = \sqrt{k^2+m^2+(\lambda/2)\phi_0^2} $ (the square root
is real for all $ k $, because $ m^2+(\lambda/2)\phi_0^2 > 0 $).
These initial conditions correspond to the quantum vacuum (no
excitations).

\bigskip

It is convenient to introduce the following dimensionless quantities
[in addition to those in Eq. (\ref{sindim})],
\begin{eqnarray}
&&  q \equiv \frac{k}{|m|} \quad ; \quad
  g \equiv \frac{\lambda}{8\pi^2} \quad ; \quad
  {\cal M} \equiv \frac{{\cal M}_d}{|m|} \quad ;  \cr \cr
&&
  \varphi_q(\tau) \equiv \sqrt{|m|} \; \varphi_k(t)  \quad ; \quad
  g\Sigma(\tau) \equiv \frac{\lambda}{2|m|^2}\;\langle\pi^2(t)\rangle
    \nlabel{gsigmadef} \quad ;
\nlabel{dimlvar}
\end{eqnarray}
where $ m $ and $ \lambda $ stand for the renormalized mass of the
inflaton and the renormalized self-coupling, respectively
\cite{higheuni,noscos}.

The quantum evolution equation in dimensionless variables are then for
$ \tau > 0 $,
\be \label{evolu}
\ddot \eta(\tau)  + {\cal M}^2(\tau)\; \eta(\tau) = - j \quad ; \quad
\ddot \varphi_q(\tau) + \omega_q^2(\tau)\;\varphi_q(\tau) = 0
\ee
with
\bea\label{evolu2}
&&{\cal M}^2(\tau) = -1 + \eta^2(\tau) + g\Sigma(\tau) \quad ; \quad
  \omega_q^2 = q^2 + {\cal M}^2(\tau) \cr \cr
&&g\Sigma(\tau) = g \int{q^2\;dq\;\left[|\varphi_q(\tau)|^2-{\cal
S}(\tau)\right]} \quad ; \quad
  {\cal S}(\tau) = \frac{1}{q} - \frac{\theta(q-1)}{2q^3}\,
  \frac{{\cal M}_d^2(\tau)}{|m|^2} \;.
\eea

And the initial conditions are
\bea\label{condin}
\eta(0) &=& \eta_0 \quad\quad\quad ; \quad\quad\quad \dot \eta(0) = 0
\cr \cr
\varphi_q(0) &=& \frac{1}{\sqrt{\omega_q(0)}}  \quad ; \quad
  \dot \varphi_q(0) = -i\; \sqrt{\omega_q(0)} \;.
\eea
where $ \eta_0 $ is given by Eq. (\ref{eta1}); and
$ \omega_q(0) = \sqrt{q^2-1+\eta_0^2} $ because the initial conditions
imply $ g\Sigma(0) = O(g) \ll 1 $. ($ \omega_q $ is real for all $ q $, 
because $ \eta_0^2 > 1 $).

It is important to note that the quantum effective mass equals the
classical effective mass plus an additional quantum term, $
g\Sigma(\tau) $.
This term comes from the in-medium effects due to the presence of
$ \vec \pi $ particles.

In this section we have exposed the quantum equations of motion and
the initial conditions using the Heisenberg picture.
The equations shown for the large $ N $ limit are the same as those
obtained
in the Schr\"odinguer picture with a gaussian {\em ansatz} for the wave
functional
(as it has been shown in Refs. \cite{renorm,equiv})
\be
\Psi[\vec\pi;\tau] = {\cal N}_\Psi \, \prod_{\vec q}
  e^{- \frac{A_{\vec q}(\tau)}{2}\, \vec\pi_{\vec q}\cdot\vec\pi_{-{\vec
q}}}
\ee
where $ {\cal N}_\Psi $ is the normalization factor.
The relation between the two pictures is made explicit by using the relation
$ A_q = -i\, \dot\varphi_q^* / \varphi_q^* $ .

$ g\Sigma(\tau) $ gives us the spreading of this wave functional,
as we see from Eq. (\ref{gsigmadef})
(the expectation values are independent of the picture).

\subsection{Early time quantum evolution}

At early times, $ g\Sigma(\tau) $ is negligible in the $ \eta $
evolution equation because $ g\Sigma(0) = O(g) \ll 1 $.
Thus, for early times the quantum dynamics of the expectation
value $ \eta(\tau) $ is the same as the dynamics of the classical
field presented in section \ref{sclasevol}. This has been explicitly
verified from the numerical solution of the full quantum equations
(\ref{evolu})--(\ref{condin}). As in the classical case, we can
distinguish two dynamical regimes depending on the value of $ j $
compared with $ j_c $.

Later on, this picture is modified due to the effects of the spinodal
 instabilities  and parametric resonances. They make the quantum modes
 to grow,  and therefore so does $ g\Sigma(\tau) $. Parametric
 resonances arrive because $ {\cal M}^2(\tau) $ oscillates in time.
In addition, the modes grow exponentially (spinodal instabilities) in
 the time intervals where  $ {\cal M}^2(\tau)  < 0 $.

These effects have a quantum nature, and they change the dynamics of the
expectation value $ \eta(\tau) $, due to its coupling with the quantum
modes
through $ g\Sigma(\tau) $. $ g\Sigma(\tau) $ gives the spreading of
the field wave functional in the transverse directions as shown by
Eq. (\ref{gsigmadef}) (recall that the transverse fields have zero
expectation value).

We comment below on the effects of this phenomena in the dynamics.

\subsubsection{$ j > j_c $}

For early times the dynamics of the field expectation value is the
same as in the classical case. Thus, the system in this case has
enough energy to jump  above the barrier and reach the region of
the global minimum of the potential (see Fig. \ref{fclaspotgjc}).
The system continues
to jump  above the barrier back and forth until the effects of the
quanta
created through spinodal instability and parametric resonance become
important, i.e., $ g\Sigma(\tau) \sim 1 $
(see Figs. \ref{fetagjc}--\ref{fm2gjc}).

The system is spinodally instable when $ {\cal M}^2(\tau) < 0 $, i.e.,
for
$ \eta^2(\tau) + g\Sigma(\tau) < 1 $, that for early times ($ g\Sigma
\simeq 0 $)  reduces to $ |\eta(\tau)| < 1 $. It also has parametric
resonances due to the oscillations of  $ {\cal M}^2(\tau) $. Both
mechanisms lead to an abundant  creation of quanta transferring
energy from the $ \eta
$ degree of freedom to the quantum fluctuations $ \varphi_k(\tau)
$. Therefore, $ g\Sigma(\tau) $ increases while
the amplitude of the  oscillations of $ \eta(\tau) $ decreases.
See Figs. \ref{fetagjc} and \ref{fgsgjc}.

\subsubsection{$ j < j_c $}

The early dynamics is the same as in the classical treatment.
For $ j < j_c $ the system does not have enough energy to jump  above
the
barrier and reach the global minimum, and it oscillates
around the metastable minimum (false vacuum). (See Fig.
\ref{fclaspotljc}).

The field could go classically to the global minimum region
moving in the transverse directions. But this possibility is excluded
due to the initial conditions Eq. (\ref{condit}): the transverse field
components and its derivatives are zero and they remain zero by the
classical evolution.  See section \ref{sclasevol}.

In the quantum treatment there are in general other two  ways of
reaching the global minimum region: one is tunneling through the
barrier and the 
second is allowing to circumvent the maxima by increasing the quantum
probability of finding large values of the transverse field components.
The tunneling is suppressed in the large $ N $ limit
because it is related to the quantum effects in the longitudinal
directions which are subleading  for large $ N $ (see comments in
section \ref{squantumeq}).
In the large $ N $ limit the leading quantum effects come from the
transversal field components and those are which we describe.

In the early (classical like) dynamics, $ \eta(\tau) $ oscillates
between
$ \eta_0 $ and $ \eta_t $ given by Eq. (\ref{clasturnp})
(see Fig. \ref{fetaljc}).
Thus, as $ g\Sigma(\tau) = 0 $ the effective mass oscillates between
the values (see Figs. \ref{fgsljc}, \ref{fm2ljc}),
\bea
{\cal M}_{max}^2 &=& -1 + \eta_0^2 = j - \frac12 j^2 + \frac58 j^3 - j^4
  + O(j^5)  > 0 \\
{\cal M}_{min}^2 &=& -1 + \eta_t^2 = - 3j - \frac12 j^2 - \frac{23}{8}
j^3
  - 4 j^4 + O(j^5) < 0
\eea
Therefore, the low momentum modes grow exponentially in the time
intervals where the effective mass is negative.
This spinodal instability increases $ g\Sigma(\tau) $ yielding a
non-zero
probability for the transverse field components.

These instabilities are well approximated by  considering a constant
negative
squared mass $ -\mu^2 $ given by the average
\be
-\mu^2 \simeq \frac{ {\cal M}_{min}^2 + {\cal M}_{max}^2 }{2} =
  -j - \frac12 j^2 - \frac98 j^3 - \frac52 j^4 + O(j^5)
\ee
This approximation reproduces the growth of $ g\Sigma(\tau) $ in good
agreement with the full numerical solution of Eqs.
(\ref{evolu})--(\ref{condin})
(see appendix \ref{apspin})
\be
g\Sigma(\tau) \approx \frac{g\,\sqrt{\mu \,\pi}}{8}\;
\frac{e^{2\tau\mu}}{\tau^{3/2}} \; .
\ee
Thus the probability of finding large values for the transverse field
components increases with time.

This solution  holds till a time $ \tau_s $ (spinodal time)
when $ g\Sigma(\tau_s) \sim \mu^2 $, this condition gives $ \tau_s $ as
the  solution of
\be
\tau_s \approx \frac{1}{2\mu}\, \log\left(\frac{8}{g\,\sqrt{\pi}}\right)
+ \frac{3}{4\mu}\log(\mu\tau_s) = \frac{1}{2\mu}\,
\log\left(\frac{8}{g\,\sqrt{\pi}}\right)+   {3 \over 4 \mu
}\log\left[\frac12 \log\left(\frac{8}{g\,\sqrt{\pi}}\right)\right] +
{\cal O}\left({ \log\log{1 \over g} \over \log{1 \over g}}\right) \;.
\ee
After $ \tau_s $ the  effect of $ g\Sigma(\tau) $ in
the evolution equations become crucial.

Soon after  $ \tau = \tau_s $ the order parameter  $ \eta(\tau) $ makes
a
spectacular jump and starts to oscillate near the global minimum
[see Figs. \ref{fclaspotljc}, \ref{fetaljc}].
This transition is explained by the fact that $ g\Sigma(\tau)=
\frac{\lambda}{2|m|^2}\;\langle\pi^2(t)\rangle_R  \sim 1
$ implies  transverse field components of the order $
|m|/\sqrt{\lambda} $. These large transverse field fluctuations allow
the system to turn around the maximum at
$ \vec{\Phi} = (\sqrt{N}\phi,\, \vec\phi_\pi) =
( |m|\sqrt{2 \over\lambda} \; \eta_{M'} \; , \vec 0) $ and reach the
region
around the global minimum where it oscillates quasi-periodically (see
Figs.  \ref{fclaspotljc}, \ref{fetaljc}--\ref{fcompana}).

The growth of  $ g\Sigma(\tau) $ allows $ \eta(\tau) $ to reach the
region around the global minimum. That is, the  growth of the
transverse quantum fluctuations elliminate the barrier that keeps $
\eta(\tau) $ near the metastable minimun.

\subsection{Intermediate time quantum evolution}

After the early period described in the previous section,
the system enters in a quasi-periodic regime. Recall that for
different initial conditions the oscillations damped much faster
\cite{j0eta, tsunos}.

The quasi-periodic regime behaviour found for  $ \eta(\tau) $ and $
g\Sigma(\tau) $ suggests that these quantities are approximately
governed by an effective hamiltonian with a few degrees of
freedom. Actually, we observe from the full numerical solution of
Eqs. (\ref{evolu})--(\ref{condin}) that $  g\Sigma(\tau) $ and  $
\eta(\tau) $ turn to be  approximately related as
\be\label{Sigeta}
g\Sigma(\tau) = 1 + \alpha \, j - (1 - \beta \, j)[\eta(\tau) +j]^2
\ee
where $ \alpha $ and $ \beta $ are positive numbers of the order $ j^0
$ and $ g^0 $ for small $ g $. The  coefficients $ \alpha  $ and $
\beta $ are obtained by fitting to the numerical solution. See   Table
\ref{tc0c1}.

\bigskip

\begin{table}[h]
\begin{tabular}{|@{~}c@{~}||@{~}c@{~}|@{~}c@{~}|@{~}c@{~}|@{~}c@{~}|@{~}c@{~}|@{~}c@{~}|@{~}c@{~}|}

\hline
$ j   $ & $ 0.05 $ & $ 0.10 $ & $ 0.15 $ & $ 0.20 $ & $ 0.24 $ &
$ 0.25 $ & $ 0.30 $ \\ \hline \hline
$ \alpha $ & $ 0.74 $ & $ 0.67 $ & $ 0.60 $ & $ 0.56 $ & $ 0.35 $ &
$ 0.29 $ & $ 0.18 $ \\ \hline
$ \beta  $ & $ 1.36 $ & $ 0.91 $ & $ 0.60 $ & $ 0.40 $ & $ 0.42 $ &
$ 0.49 $ & $ 0.43 $ \\ \hline
\end{tabular}
\caption{ $ \alpha $ and $ \beta $ values obtained fitting $
g\Sigma(\eta) $
in the time interval $ \tau \in [800,\,1000] $ for $ g = 10^{-6} $ and
various $ j $ values.  \label{tc0c1}}
\end{table}

\bigskip

We thus find for the effective  squared mass,
\be  \label{M2etarel}
{\cal M}^2(\eta) = -1 + \eta^2 + g\Sigma(\eta)
= j \left[ \alpha - j +\beta j^2 -2( 1 - \beta j )\, \eta +\beta \,
\eta^2 \right]
\ee

The evolution equations (\ref{evolu})--(\ref{evolu2}) in this
approximation take then the  form,
\be
\ddot \eta + {\cal M}^2(\eta)\, \eta = - j
\ee

We find integrating on $ \eta $,
\be\label{efE}
\frac12 {\dot \eta}^2 + j \; V_{int}(\eta) = j\, E_{int}
\ee
where,
\bea
V_{int}(\eta) &=& \eta + \frac12 (\alpha - j +\beta j^2)\eta^2
-\frac23( 1 - \beta j )\, \eta^3
+{\beta \over 4} \,\eta^4  \:, \\
E_{int} &=& V_{int}(\eta_1)
\eea
Notice that  $E_{int}$ depends on the initial conditions,
and that $ \eta_1 $ is a turning point of the motion.
Eq. (\ref{efE}) can be integrated as follows,
\be\label{intel}
\sqrt{2 \, j}\,(\tau-\tau_1) = \int_{\eta_1}^\eta{
\frac{d\eta}{\sqrt{E_{int} -  V_{int}(\eta)}}}
\ee
with $ \eta_1 = \eta(\tau_1) $.

The fourth order polynomial $ E_{int} -  V_{int}(\eta) $ has always
two real roots $ \eta_1 < \eta_2 $ corresponding to the turning
points. Depending on the value of $ j $, the two other roots are a
complex conjugated pair or two more real roots. For each case:

\begin{itemize}
\item
i) a pair of complex conjugate roots $ \eta_R \pm i \eta_I $. We then
define:
\be
a \equiv \frac{\eta_R-\eta_1}{(\eta_R-\eta_1)^2+\eta_I^2} \; ; \quad
b \equiv \frac{\eta_I}{(\eta_R-\eta_1)^2+\eta_I^2} \; .
\ee
\item
ii) a pair of real roots $ \eta_1 < \eta_2 < \eta_3 < \eta_4 $. We
then define:
\be
a \equiv \frac{2}{\eta_3-\eta_1} - \frac{1}{\eta_4-\eta_1} \; ; \quad
b^2 \equiv {4 (\eta_3-\eta_2)(\eta_4-\eta_3) \over
(\eta_2-\eta_1)(\eta_4-\eta_1) (\eta_3-\eta_1)^2}\; .
\ee
\end{itemize}
We also introduce two  other quantities to simplify the formulae:
\be\label{cosas}
d \equiv \frac{1}{\eta_2-\eta_1} \; ; \quad
X \equiv \left[ (a-d)^2 + b^2 \right]^{1/4} \; ; \quad
C \equiv X\sqrt{2|{\cal M}^2(\eta_1)\,\eta_1+j|}
\ee
It is convenient to use in Eq. (\ref{intel}) as integration variable
$ u \equiv {1 \over \eta-\eta_1}$.

The solution of Eq. (\ref{intel}) can be expressed after calculation as
\be  \label{etaanasol}
\eta(\tau) = \eta_1 +
\frac{(\eta_2-\eta_1)[1-\mbox{cn}(C(\tau-\tau_1),k)]}{1+(\eta_2-\eta_1)X^2
-[1-(\eta_2-\eta_1)X^2]\; \mbox{cn}(C(\tau-\tau_1),k)} \; ,
\ee
where $ \mbox{cn}(z,k) $ is the Jacobi cosine function, and
\be
k = \frac{1}{\sqrt{2}} \sqrt{1+\frac{a-d}{X^2}}
\ee
the elliptic modulus.

The solution (\ref{etaanasol}) oscillates between $ \eta_1 $ and $
\eta_2 $
with period
\be
T = \frac{2}{C}\, K(k)
\ee
where $ K(k) $ stands for the complete elliptic integral of the
first kind and $ C $ is given by Eq. (\ref{cosas}).
The analytical and the numerical solutions are compared in Fig.
\ref{fcompana}.

The numerical solution would be exactly periodic if $ E_{int} $ were
exactly
conserved; however it is slowly decreasing.
The damping is small the smaller is $ j $. See Table \ref{tEint}.

\bigskip

\begin{table}[h]
\begin{tabular}{|@{~}c@{~}||@{~}c@{~}|@{~}c@{~}|@{~}c@{~}|@{~}c@{~}|}
\hline
$ \tau $             & $500$  & $1000$  & $1500$  & $2000$  \\  \hline
\hline
$ j \, E_{int} (j=0.05) $ & $0.02$ & $0.016$ & $0.014$ & $0.011$ \\
\hline
$ j \, E_{int} (j=0.20) $ & $0.04$ & $0.025$ & $0.015$ & $0.010$ \\
\hline
\end{tabular}
\caption{ $ E_{int} $ for $ g = 10^{-6} $ and $ j = 0.05 $ and $ 0.20 $.
\label{tEint}}
\end{table}

\bigskip

Thus, $ E_{int} $ slowly decreases while $ \eta(\tau) $ and $ {\cal
M}^2(\eta) $ oscillate fast showing a clear separation between slow
and fast variables. We have explicitly found the fast time dependence
of the order parameter and the quantum fluctuations in terms of
elliptic functions. Comparison with the full numerical solution shows
that Eqs. (\ref{Sigeta}) and (\ref{etaanasol}) actually
reproduce the full dynamics provided the constant parameters in
Eq. (\ref{etaanasol}) become {\bf slow} functions of time. For example, the
parameters displayed in Table \ref{tc0c1} change by $ 3 - 4 \% $
between $ \tau = 500 $ and $ \tau = 1000 $. The study of such
slow dynamics is a very interesting problem beyond the scope of this
article and that can be solved using Whitham methods \cite{whi}.

As is known, for $ j = 0 $ the dynamics is governed by a exactly flat
effective potential for $ |\eta | < 1 $ \cite{j0eta}. Since the slow
variables become here slower for $ j \to 0 $, they may be governed
 by an effective potential becoming flat at $ j = 0 $.

Now, we can obtain analytic expressions for $ g\Sigma(\tau) $ and
$ {\cal M}^2(\tau) $ from the relations $ g\Sigma(\eta) $
[Eq. (\ref{Sigeta})] and $ {\cal M}^2(\eta) $  [Eq. (\ref{M2etarel})],
just using the analytical solution $ \eta(\tau) $ [Eq.
(\ref{etaanasol})].
The relation $ g\Sigma(\eta) $ indicates that $ g\Sigma $ and $ \eta $
oscillate with opposite phase, up to order $ j $ terms;
$ g\Sigma(\eta) = 1 - \eta^2 + O(j) $. This $ O(j) $ terms are very
important
because  the $ j^0 $ terms cancel in the effective mass, and the $ j $
terms
make the mass to be different from zero.
This has the interesting result of making the time averaged squared mass
positive. Such average tends asymptotically to a positive value.
[Recall that for $ j=0 $ in the broken symmetry case the squared mass
goes to zero for initial energies below the potential at the local
maximum ($\eta = 0 $) \cite{j0eta,tsunos}.]

The effective mass vanishes asymptotically for $ j \to 0 $.
This can be seen explicitly from Eq. (\ref{M2etarel});
it implies  $ {\cal M}^2(\tau = \infty) = {\cal O}(j) $ for small $ j $.

Notice also that in the $ j=0 $ case the effective mass vanishes
asymptotically \cite{j0eta} giving the constraint
$$
1 = \eta^2 +  g\Sigma \quad \mbox{for} \quad j=0 \; ;
$$
and the thick curve in Fig. \ref{fetagsljc} is $ {\cal O} (j) $
away from such a parabola; showing that the squared mass is of order $j$.

The  mass of the scalars {\bf around the ground state} $ {\hat m} $ follows
from the effective potential [see Appendix B]. 
For $ j = 0 $ one finds that the mass $ {\hat m} $ vanishes
\cite{otros, j0eta} in accordance with the 
presence of  $N-1$ Goldstone bosons. For $ j \neq 0 $ the mass  $
{\hat m}^2 $ becomes non-zero and for small $ j $ we find in Appendix
B, 
$$
 {\hat m}^2 = j +{\cal O}(j^2 \, \log j ) \quad
 \mbox{for~the~ground~state} .
$$

The asymptotic mass squared $ {\cal M}^2(\infty) $ found above
does not coincide with $  {\hat m}^2 $ although both masses are of the same
order of magnitude $ {\cal O}(j) $.  Notice that $  {\hat m}^2 $ describes
excitations around the ground state, while $ {\cal M}^2(\infty) $
corresponds to excitations around states with finite energy density
above the ground state.

\subsection{Constraints on the trajectory due to the conservation of the
energy}

As the states considered here are homogeneous and isotropic, the system
has an energy-momentum tensor with the ideal fluid form.
Thus, we can define a dimensionless energy as
\bea \label{defe}
\epsilon &\equiv& \frac{\lambda}{2N|m|^4}\,\langle T^{00} \rangle \cr\cr
&=& -\frac{\eta^2}{2} + \frac{\eta^4}{4} +\frac12\,\eta^2 g\Sigma
  - \frac{g\Sigma}{2} +\frac{(g\Sigma)^2}{4}+\frac14 + j\eta \cr\cr
&&+ \frac{\dot\eta^2}{2} +\frac{g}{2}\int{q^2dq\,|\dot\varphi_q|^2}
  + \frac{g}{2}\int{q^2dq\,q^2\,|\varphi_q|^2} \; .
\eea
for $ \tau > 0 $. 

The conservation of the energy (for constant external fields)
gives us the following constraint in the $ (\eta,\,g\Sigma) $ plane
\begin{equation} \label{econs}
\epsilon_0 \ge V_{eff \; dyn}(\eta,\Sigma)\equiv  -\frac{\eta^2}{2} +
\frac{\eta^4}{4} + \frac12\,\eta^2 g\Sigma -\frac{g\Sigma}{2} +
\frac{(g\Sigma)^2}{4} + \frac14 + j\eta \; .
\end{equation}
where $ \epsilon_0 = V_{\tau \leq 0}(\eta_0) + 2j\eta_0 =
V_{\tau>0}(\eta_0) $
is the energy after the sign change in the external field $j$.

This constraint is represented as a dotted line in Fig. \ref{fetagsljc}.
We see there how the energy conservation constrains the trajectories in
the $ (\eta, \, g\Sigma) $ plane.

$ V_{eff \; dyn}(\eta,\Sigma) $ can be interpreted as a dynamical
effective potential for the longitudinal field $ \eta $. Notice that
the equation of motion (\ref{evolu}) for $ \eta(\tau) $ can be written as 
$$
\ddot \eta(\tau) = - {\partial \over \partial\eta}V_{eff \;
dyn}(\eta,\Sigma) \; . 
$$

$ V_{eff \; dyn}(\eta,\Sigma) $ defined by  eq. (\ref{econs})  is
plotted for $ j = 0.20 $ in Fig. \ref{fsemimex}.

\section{Comments and Conclusions}

Here, we have studied effects of external fields in
non-perturbative quantum field dynamics in the large $ N $ limit.
The effects of uniform external fields in $ \Phi^4 $ theory with
broken symmetry were recently studied in a different framework through
classical evolution with random initial conditions \cite{patkos}.

We have studied the evolution of the ground state in an external
uniform field $ J $ after flipping its sign: $ J \to -J $.
We have considered the broken symmetry case and small external
fields (then the classical potential presents two local minima).

The change of sign in the external field leads to spinodal instabilities
and parametric resonances yielding abundant particle creation of the
order $ 1/\lambda $. As a consequence, the wave  functional spreads
allowing large values for field components orthogonal to $ \vec{\cal J}
$. Thanks to such large transverse field fluctuations the system goes
around the maximum of the potential and reaches the global minimum
without tunneling.  

The large transverse fluctuations $ \Sigma(\tau) $ change the
effective mass squared in the evolution of the longitudinal field $
\eta(\tau) $ [see eqs.(\ref{evolu})-(\ref{evolu2})]. In this way $
\eta(\tau) $ overcomes the classical barrier and reaches the absolute
minimun. The effective dynamical potential $ V_{eff \;
dyn}(\eta,\Sigma) $ introduced in eq.(\ref{econs}) helps to understand
this phenomenon.  As shown in fig. 12, the barrier in the $ \eta $
direction between the two minima of $ V_{eff \; dyn}(\eta,\Sigma=0) $
dissapears in  $ V_{eff \; dyn}(\eta,\Sigma) $ for sufficiently large
$ \Sigma $. 

After that, the system oscillates reaching an almost periodic
regime. That is, fast variables (whose time evolution we explicitly
solve) oscillate periodically while other variables change slowly due
to a tiny damping. 

In this quasi-periodic regime the effective squared mass is of order $
J $. Thus, in the $ J \to 0 $ limit we recover a zero effective squared
mass consistently with the known results for $ J = 0 $ where out of
equilibrium Goldstone bosons appear \cite{j0eta,tsunos}.

The damping of the fast oscillations is here much slower than for the 
oscillations found in refs. \cite{j0eta,tsunos} for other initial 
conditions.
There, the oscillations of $ {\cal M}^2(t) - {\cal M}^2(\infty) $ 
were damped as $ 1/t $, while the damping is significatively slower
for the quasi-periodic solutions found here.

\section{Acknowledgements}

We thank Andras Patk\'os for useful discussions before this work was
started and  Daniel Boyanovsky for useful comments.

\section{Appendix}

\subsection{Appendix A: Spinodal instability} \nlabel{apspin}

Here, we obtain the early time solution for the modes in the spinodally
resonant band, and we estimate the spinodal time, $ \tau_s $.

Before entering on the calculation of $ \tau_s $, let us recall that the
the contribution of the spinodal band can be estimated using an average
mass
for the evolution. The numerical calculations show that this estimation
is
correct, until such contribution becomes of order one.

Hence,  an approximate equation for the modes reads,
\begin{equation}\label{resomo}
\left(\frac{d^2}{d\tau^2}+q^2-\mu^2\right)\varphi_q(\tau)=0 \; .
\end{equation}
The modes with $ q $ in the interval between $ 0 $ and $ \mu $ are
spinodally resonant.
There are no particles in the spinodally resonant band. Therefore the
initial
condition for modes in the resonant band are:
\begin{equation}
\varphi_q(0) = \frac{1}{\sqrt \Omega_q}
= \left( q^2 + |{\cal M}^2(0)| \right)^{-1/4},
\;\;\;\; \dot\varphi_q(0) = -i \sqrt \Omega_q
= -i \left( q^2 + |{\cal M}^2(0)| \right)^{1/4}.
\end{equation}
The solution of Eq. (\ref{resomo}) for these modes is:
\begin{eqnarray}
\varphi_q(\tau) = \frac{1}{2\sqrt{\mu^2-q^2}(q^2+|{\cal M}^2(0)|)^{1/4}}
\left[ \left(\sqrt{\mu^2-q^2}-i\sqrt{q^2+|{\cal M}^2(0)|} \right)
e^{\tau\sqrt{\mu^2-q^2}}+ \right. \nonumber \\
\left. + \left(\sqrt{\mu^2-q^2}+i\sqrt{q^2+|{\cal M}^2(0)|}\right)
e^{-\tau\sqrt{\mu^2-q^2}}\right] \;.  \label{cuadmo}
\end{eqnarray}
We thus obtain for the squared modulus neglecting
the exponentially decreasing term
\begin{eqnarray}
|\varphi_q(\tau)|^2 &\approx&
\frac{\mu^2+|{\cal M}^2(0)|}
{4\mu^2\left(1-\frac{q^2}{\mu^2}\right)\sqrt{q^2+|{\cal M}^2(0)|}}\;
e^{2\tau\mu\sqrt{1-\frac{q^2}{\mu^2}}}\; . \label{modphiapprox}
\end{eqnarray}

The contribution of the spinodal band to $ g\Sigma(\tau) $ is given by,
\begin{equation}\label{gsigspi}
\Sigma_s(\tau) = \int_0^\mu \; {q^2\; dq\; |\varphi_q(\tau)|^2} \; .
\end{equation}
Inserting Eq. (\ref{modphiapprox}) into Eq. (\ref{gsigspi}) we obtain an
estimation for the spinodal growth of the quantum fluctuations.
To approximately evaluate this integral, we can make further
simplifications
in Eq. (\ref{modphiapprox}). As $ q/\mu < 1 $ and the contribution of
the modes with $ q \approx \mu $ is exponentially suppressed,
we can expand in $ q/\mu $ to second order in the exponential and to
zeroth
order in the factor outside  the exponential.
\begin{equation}
|\varphi_q(\tau)|^2 \approx \frac{\mu^2+|{\cal M}^2(0)|}
{4\mu^2\,\sqrt{q^2+|{\cal M}^2(0)|}} \;
e^{2\tau\mu}\;e^{-\tau\frac{q^2}{\mu}} \; .
\end{equation}
In addition,  the integrand has its maximum at $ q = O(0.1\mu) $ and
$ 0 < \mu < 1 $. Therefore, we can approximate
$ \mu^2 + |{\cal M}^2(0)| \sim 2 \mu^2 $ and
$ \sqrt{q^2+|{\cal M}^2(0)|} \sim \mu $.
Thus,
\begin{equation}
|\varphi_q(\tau)|^2 \approx \frac{1}{2\mu} \;
e^{2\tau\mu}\;e^{-\tau\frac{q^2}{\mu}} \; .
\end{equation}
Then integrating over $ q $ and using $ \int_0^1 v^2 dv \exp(-v^2
r)\buildrel{r \to \infty}\over  = \sqrt{\pi/(16\,r^3)} $,
\begin{equation}
g\Sigma(\tau) \approx \frac{g\sqrt{\pi\mu}}{8}\;
\frac{e^{2\tau\mu}}{\tau^{3/2}} \; .
\end{equation}

The spinodal time $ \tau_s $ is, by definition, the time where the
instabilities are shut off for all $ 0 \leq q \leq \mu$. This happens
when the spinodal modes contribution $ g\Sigma_s(\tau) $ compensates the
initial (negative) value of  $  M^2_{eff}(\tau) = -\mu^2 $.
\begin{equation}
g\Sigma_s(\tau_s) \approx \mu^2
\end{equation}
Therefore  $ \tau_s $ is given by the following implicit equation,
\begin{equation}\label{tauese}
\tau_s = \frac{1}{2\mu}\, \log\left[\frac{8}{g\,\sqrt{\pi}}\right]
+ \frac{3}{4\mu}\log(\mu\tau_s) \;.
\end{equation}
The spinodal times given by this equation are in good agreement with the
numerical results.

\subsection{Appendix B: Ground state results}

The expectation value of the field $ \varphi $ and the mass of the scalars
can be obtained for the ground state using the effective
potential \cite{otros, j0eta}. One finds in the broken symmetry case
using dimensionless variables,
$$
\varphi^2 = 1 + {\hat m}^2 + { \lambda \, {\hat m}^2 \over  2
(4\pi)^2} \log{1  \over {\hat m}^2} \; ,
$$
where $ {\hat m} $ is the mass  around the ground state of the $N-1$ scalars.
The field expectation value is connected with the external source
by \cite{otros, j0eta} 
$$
\varphi = { j  \over {\hat m}^2} \; .
$$
Hence, $ {\hat m}^2 $ is a function of $ j $ defined by the
trascendental equation
\be\label{masngran}
{\hat m}^2 \sqrt{ 1 + {\hat m}^2 + { \lambda \, {\hat m}^2 \over  2
(4\pi)^2} \log{1  \over {\hat m}^2}} = j \; .
\ee
For $ j = 0 $ one finds that $ {\hat m}= 0 $ in accordance with the
presence of  $N-1$ Goldstone bosons. For $ j \neq 0 $ the mass  $
{\hat m}^2 $ becomes non-zero and for small $ j $ we find
$$
 {\hat m}^2 = j +{\cal O}(j^2 \, \log j ) \quad
 \mbox{for~the~ground~state}.
$$
[Notice that Eq.(\ref{masngran}) becomes the classical equation
Eq.(\ref{cubica}) for $ \lambda = 0 $ setting $  {\hat m}^2 = - 1 +
\eta^2 $, as expected.]

\begin{figure}[h]
\epsfig{file=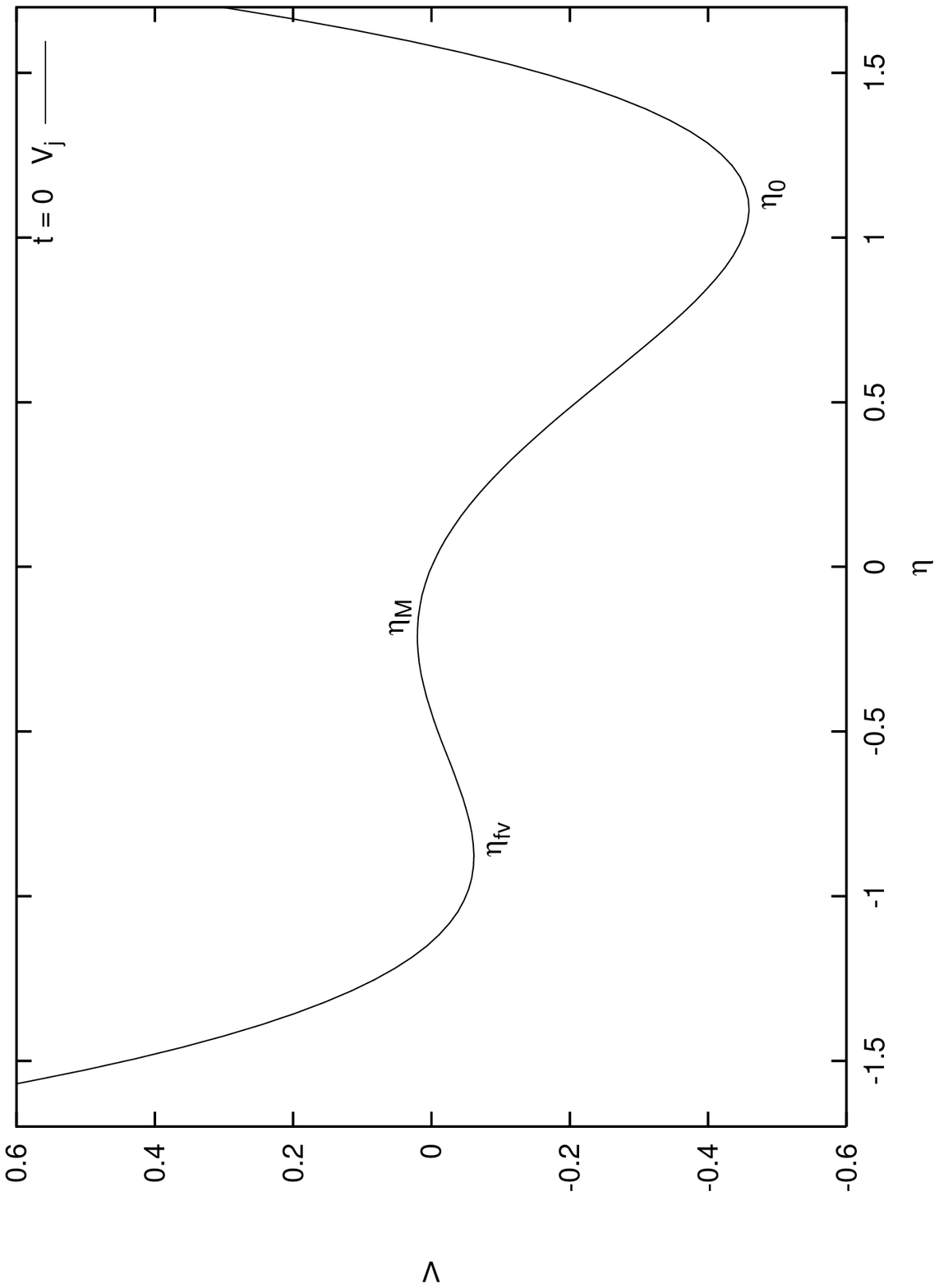}
\caption{Classical potential at $ \tau \leq 0 $ for small field
$ j = 0.20 < 2/\sqrt{27} $. The potential presents two minima (the
ground state $ \eta_0 $ and a metastable false vacuum state $ \eta_{fv} $)
separated by a potential barrier.}
\nlabel{fclaspotljcini}
\end{figure}

\begin{figure}[h]
\epsfig{file=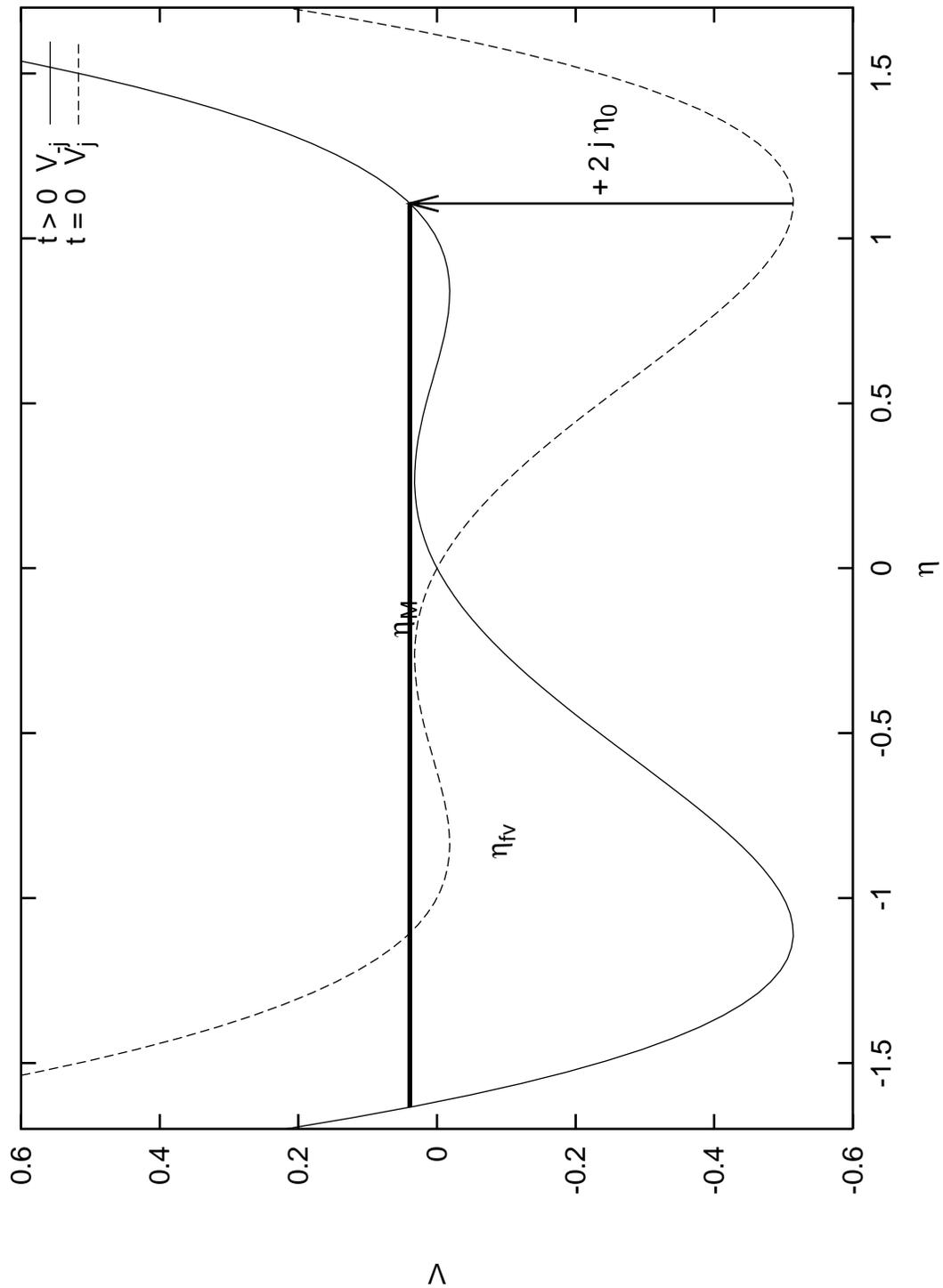}
\caption{$ j = 0.25 > j_c $. Classical potential at $ \tau = 0 $ and
for $ \tau > 0 $.
The change of the sign of the external field at $ \tau = 0 $
increases in $ 2 j \eta_0 $ the energy density of the system, and the
system has for $ j > j_c $ enough energy to jump  above the barrier.}
\nlabel{fclaspotgjc}
\end{figure}

\begin{figure}[h]
\epsfig{file=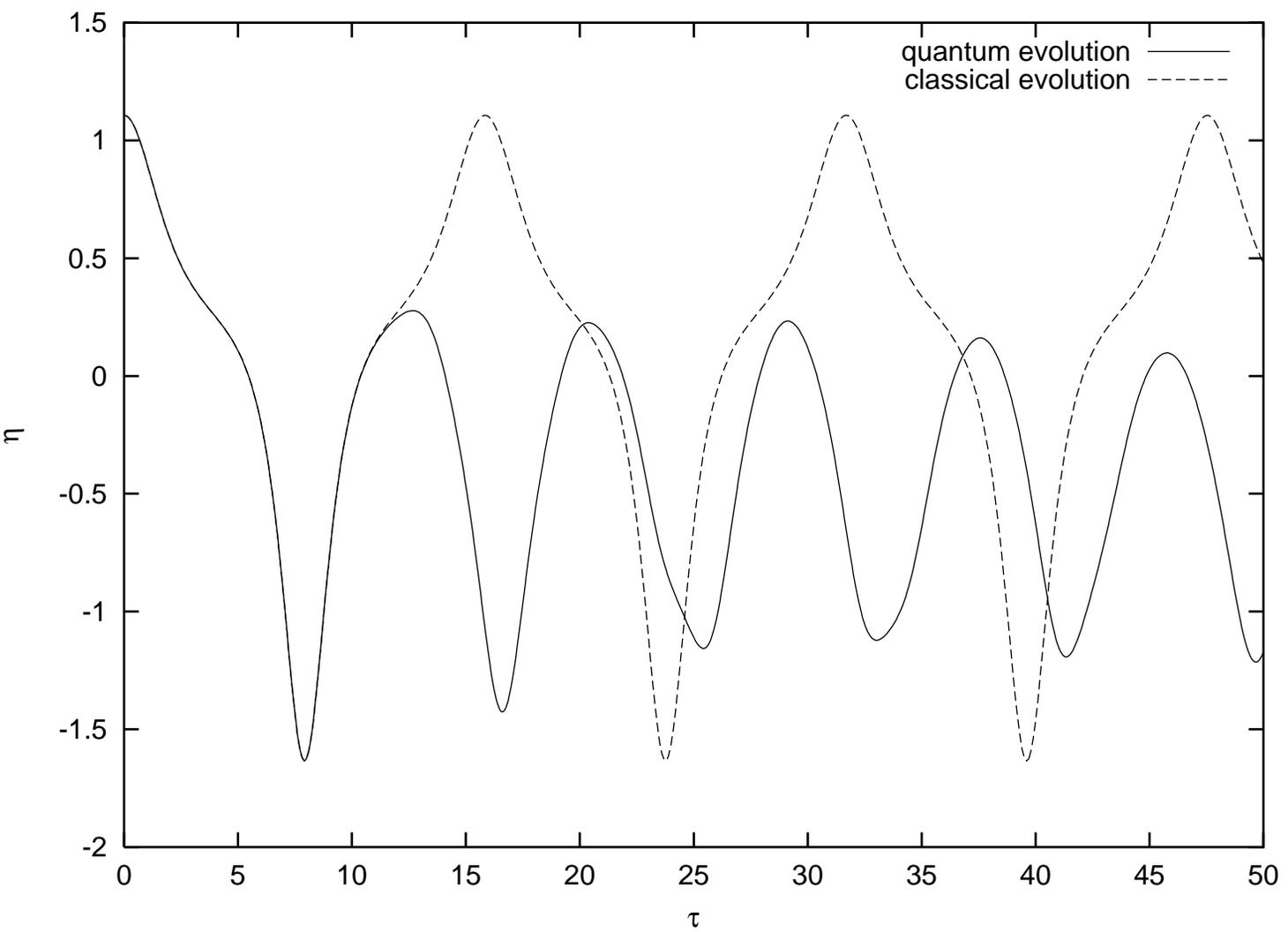}
\caption{$ j = 0.25 > j_c $.
$ \eta(\tau) $ from the quantum (full line) and the classical (dashed
line) evolution.}
\nlabel{fetagjc}
\end{figure}

\begin{figure}[h]
\epsfig{file=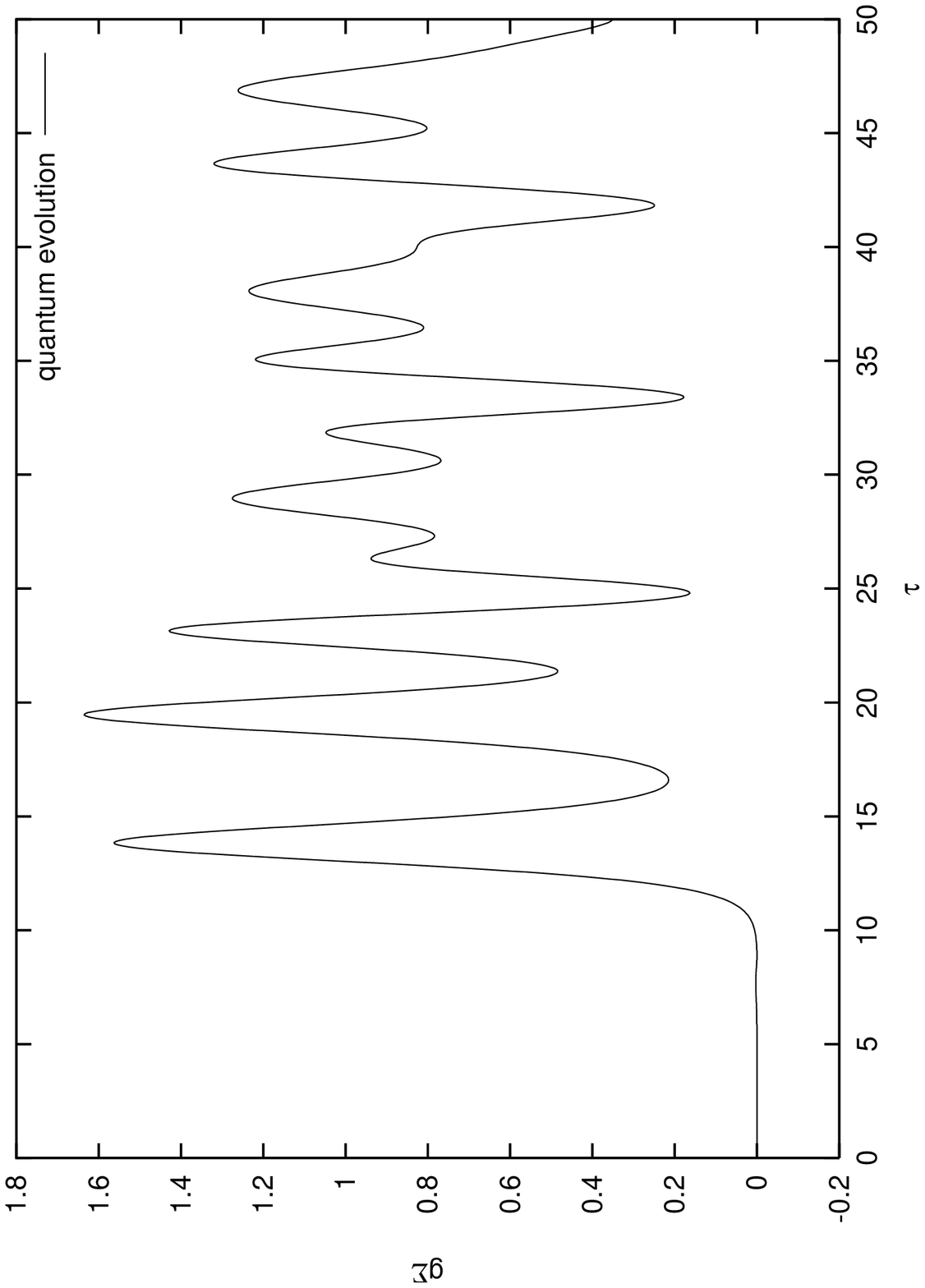}
\caption{$ j = 0.25 > j_c $.
$ g\Sigma(\tau) $.}
\nlabel{fgsgjc}
\end{figure}

\begin{figure}[h]
\epsfig{file=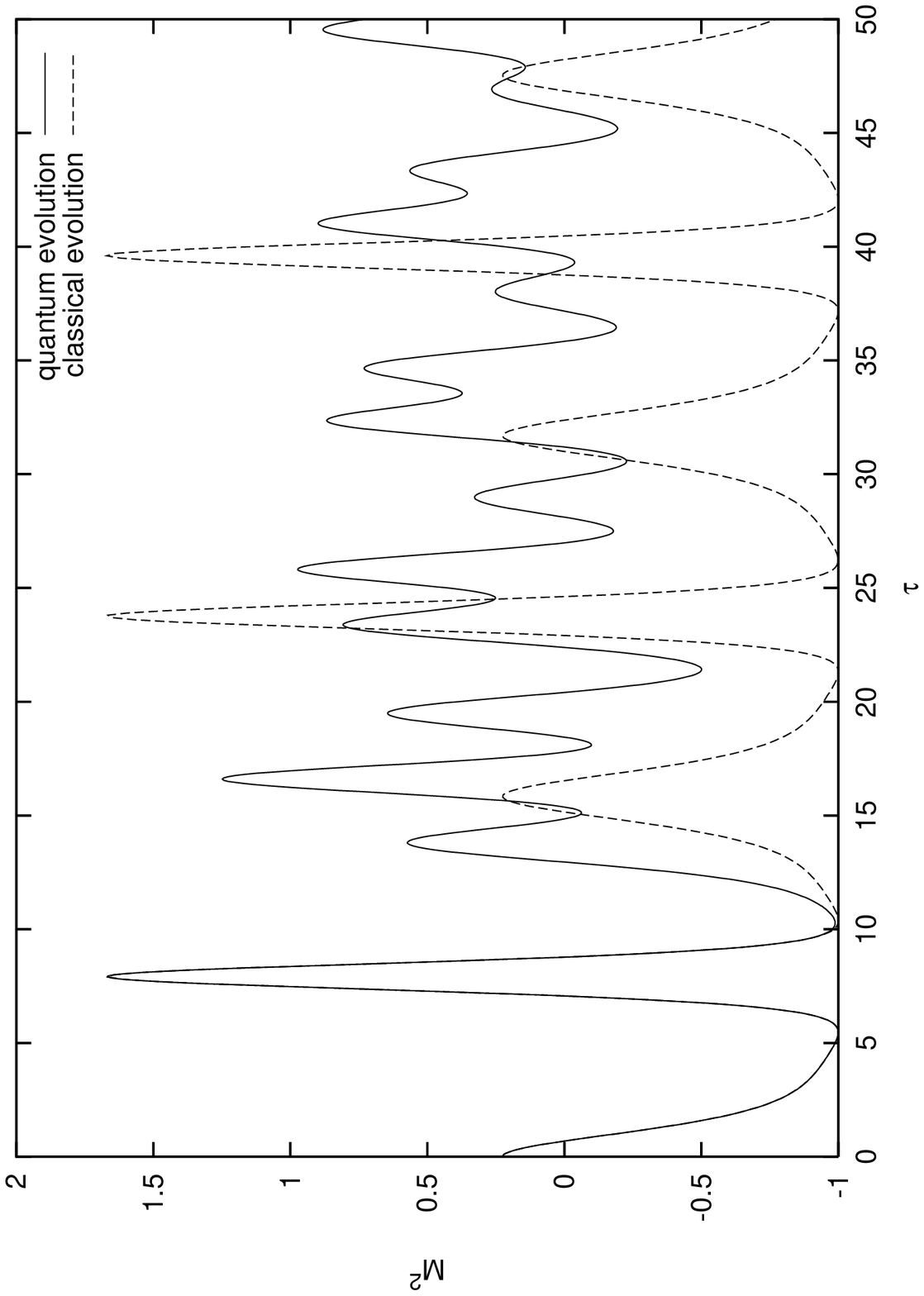}
\caption{$ j = 0.25 > j_c $.
Effective mass squared $ {\cal M}^2(\tau) $ for the quantum (full line)
and for the classical (dashed line) evolution.}
\nlabel{fm2gjc}
\end{figure}

\begin{figure}[h]
\epsfig{file=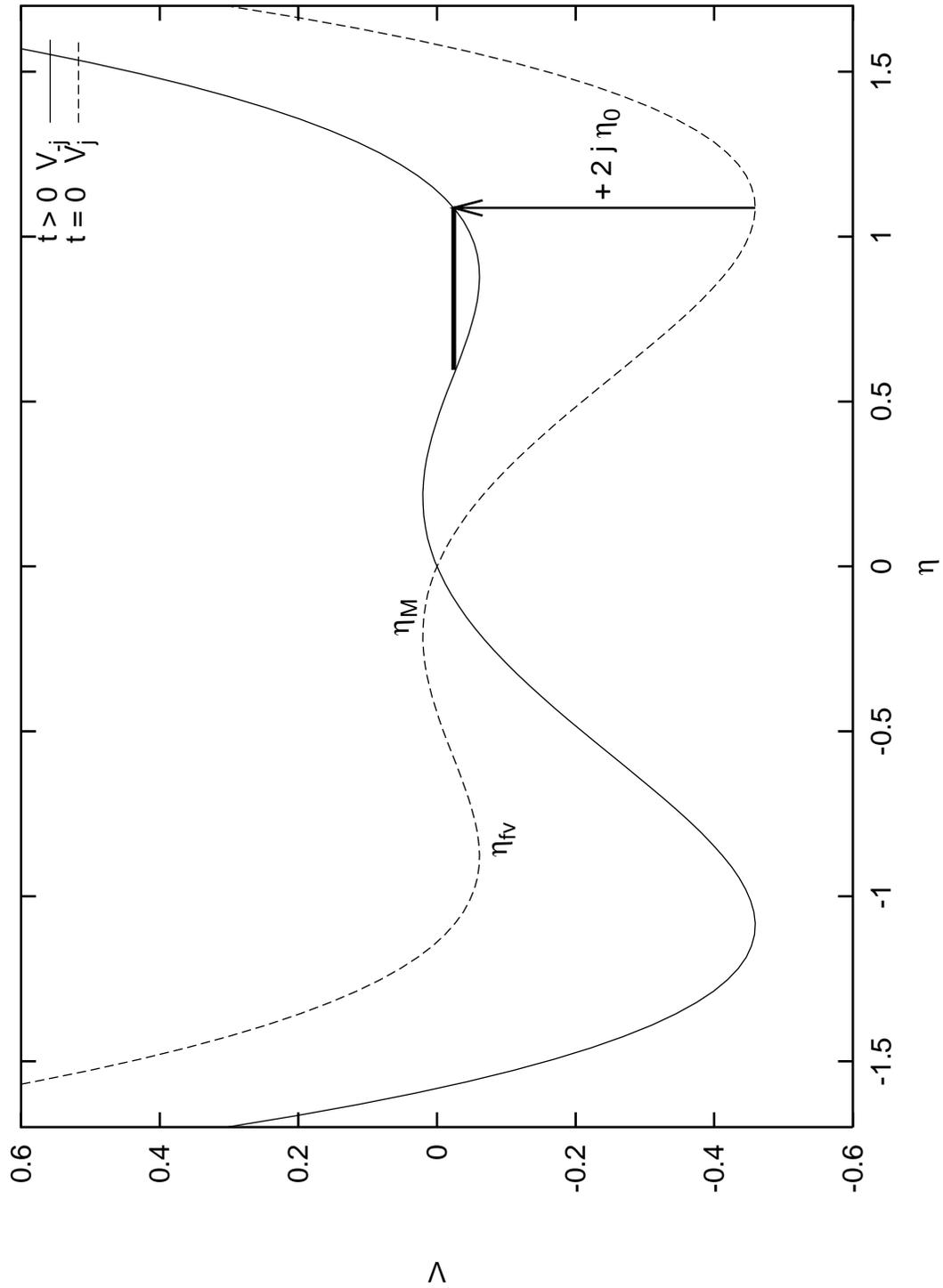}
\caption{$ j = 0.20 < j_c $. Classical potential at $ \tau = 0 $ and
for $ \tau > 0 $.
The change of the sign of the external field at $ \tau = 0 $
increases in $ 2 j \eta_0 $ the energy density of the system but
the system {\bf does not} have enough energy to jump  above the
barrier.}
\nlabel{fclaspotljc}
\end{figure}

\begin{figure}[h]
\epsfig{file=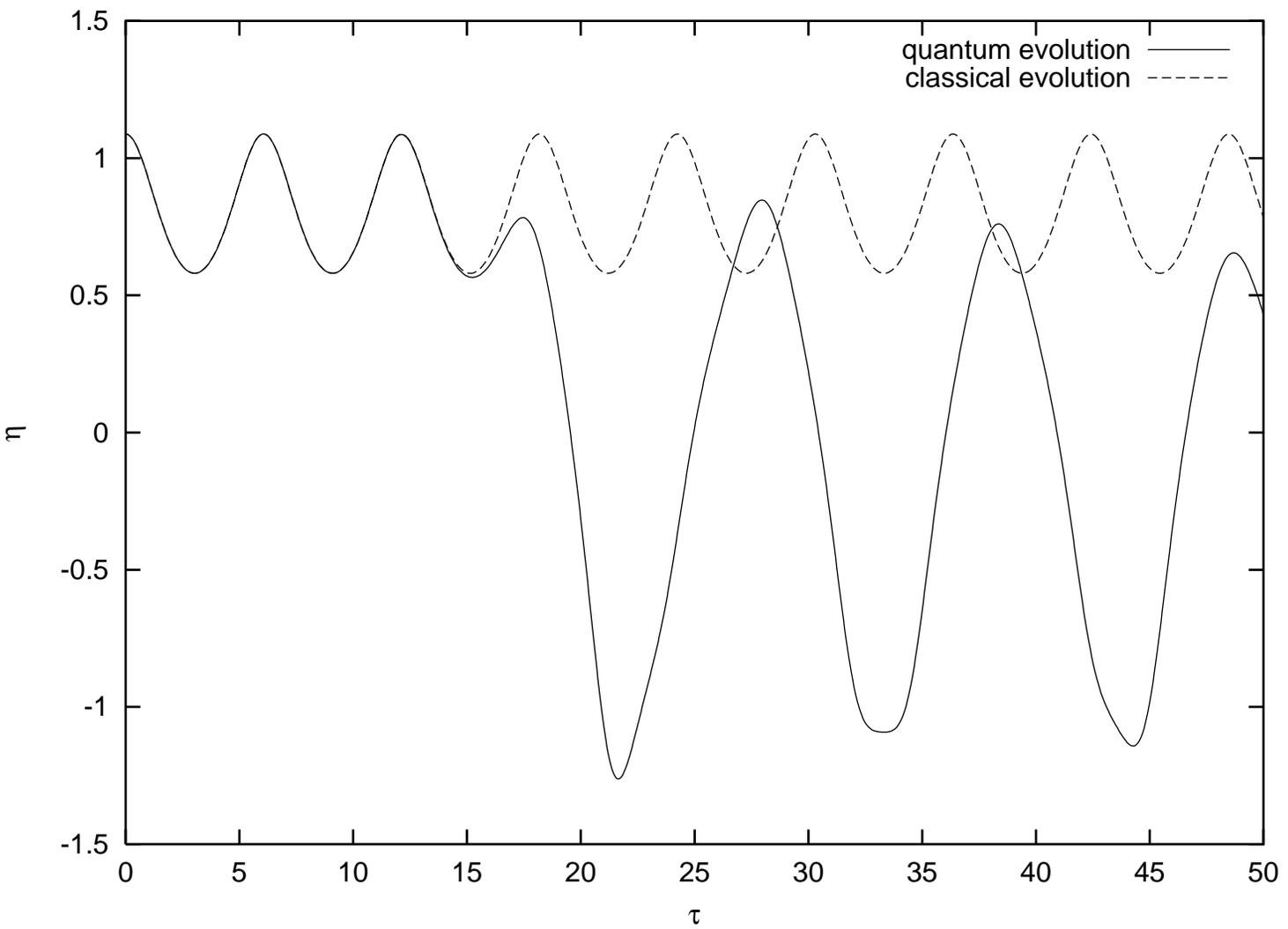}
\caption{$ j = 0.20 < j_c $. $ \eta(\tau) $ for the quantum (full line)
and for the classical (dashed line) evolution.}
\nlabel{fetaljc}
\end{figure}

\begin{figure}[h]
\epsfig{file=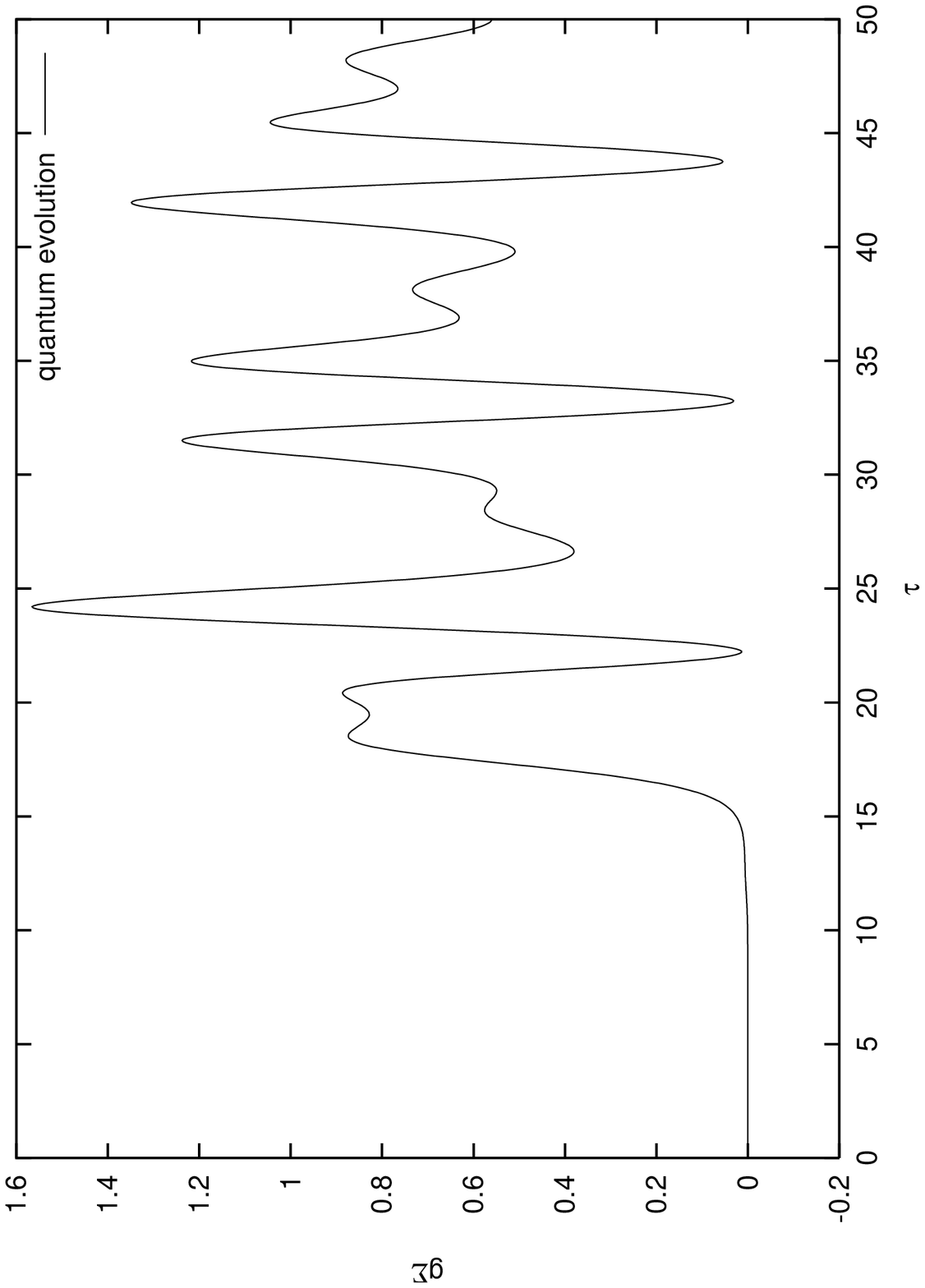}
\caption{$ j = 0.20 < j_c $. $ g\Sigma(\tau) $.}
\nlabel{fgsljc}
\end{figure}

\begin{figure}[h]
\epsfig{file=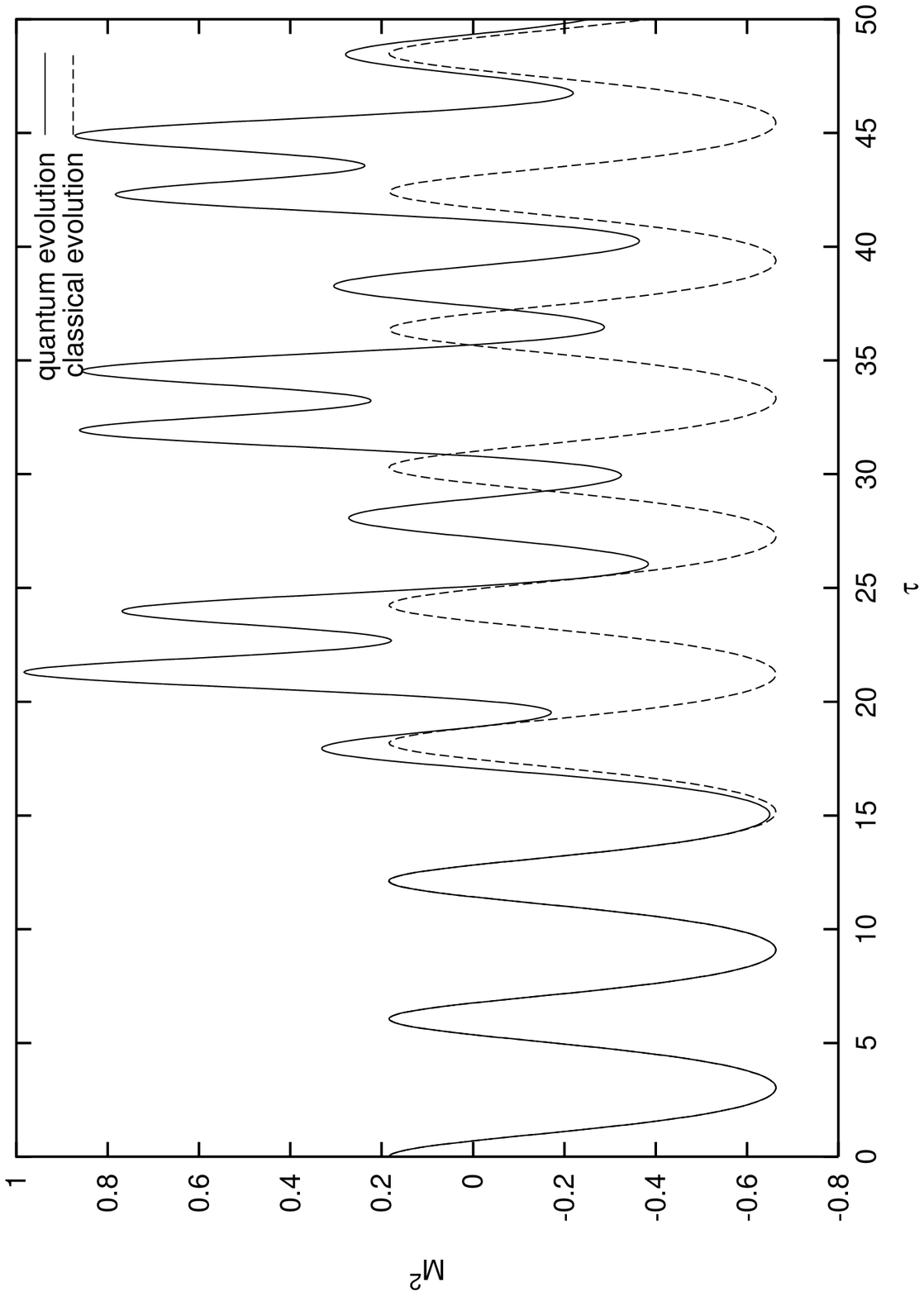}
\caption{$ j = 0.20 < j_c $.
Effective mass squared  $ {\cal M}^2(\tau) $ for the quantum (full
line) and for the classical (dashed line) evolution.}
\nlabel{fm2ljc}
\end{figure}

\begin{figure}[h]
\epsfig{file=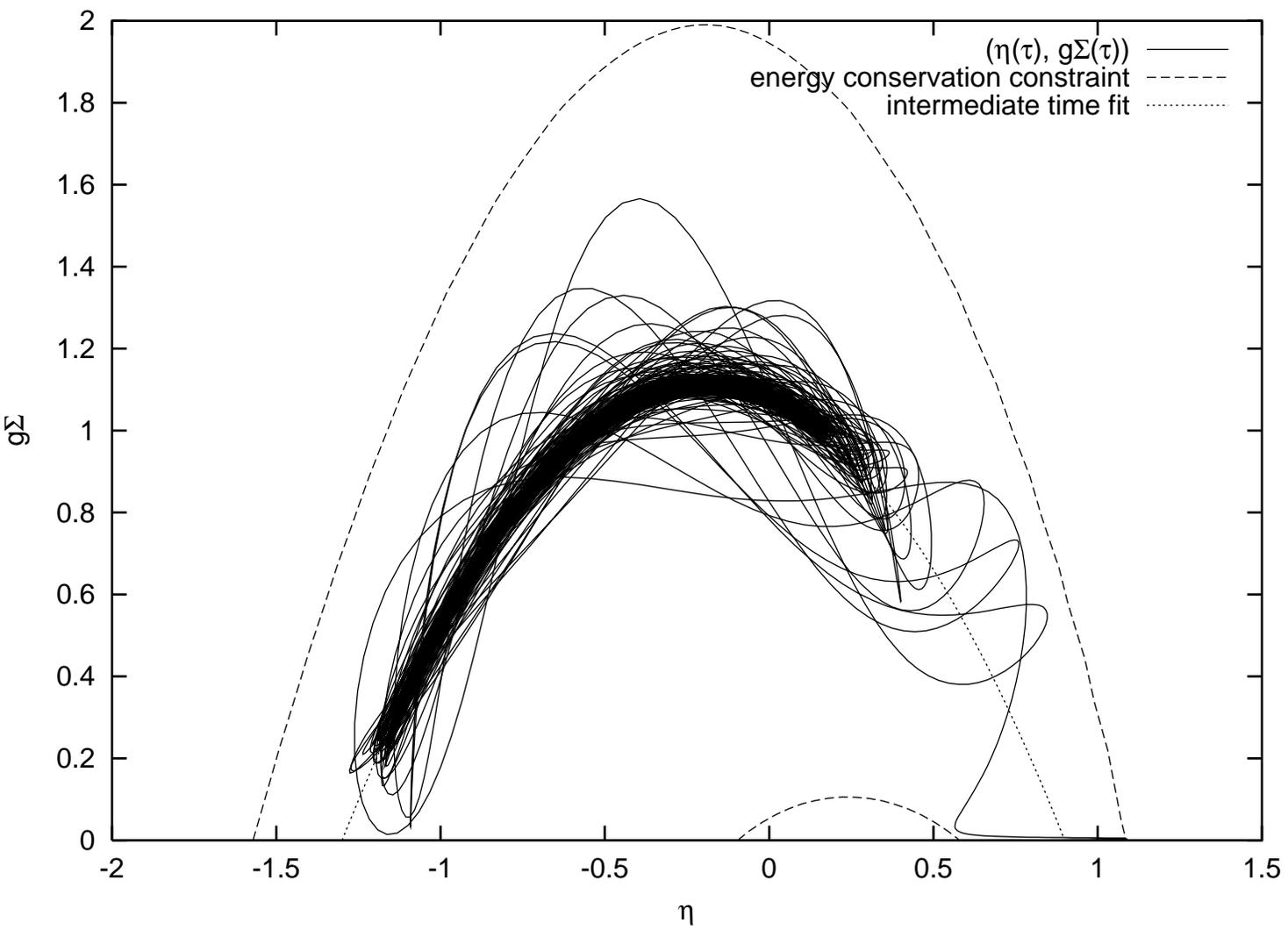}
\caption{$ j = 0.20 < j_c $. Full line: trajectory in the
$ (\eta,\,g\Sigma) $ plane. Dotted line: constraint in the trajectory
due to the energy conservation. See Eq. (\ref{econs}) and Fig.
\ref{fsemimex}.}
\nlabel{fetagsljc}
\end{figure}

\begin{figure}[h]
\epsfig{file=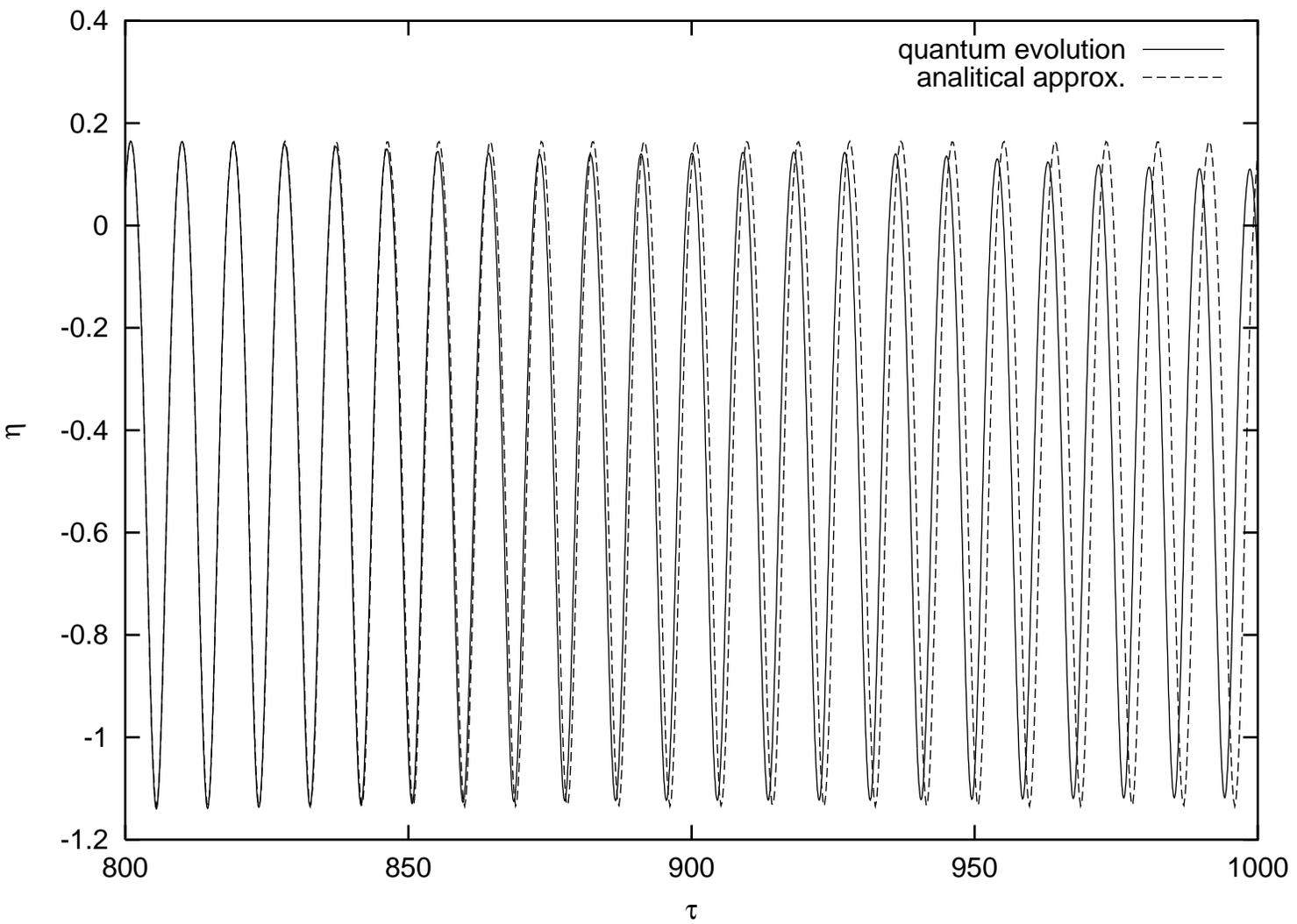}
\caption{$ j = 0.20 < j_c $. Quantum evolution for the field expectation
value $ \eta(\tau) $ (full line) compared to
the analytical intermediate time approximation (dashed line).}
\nlabel{fcompana}
\end{figure}

\begin{figure}[h]
\epsfig{file=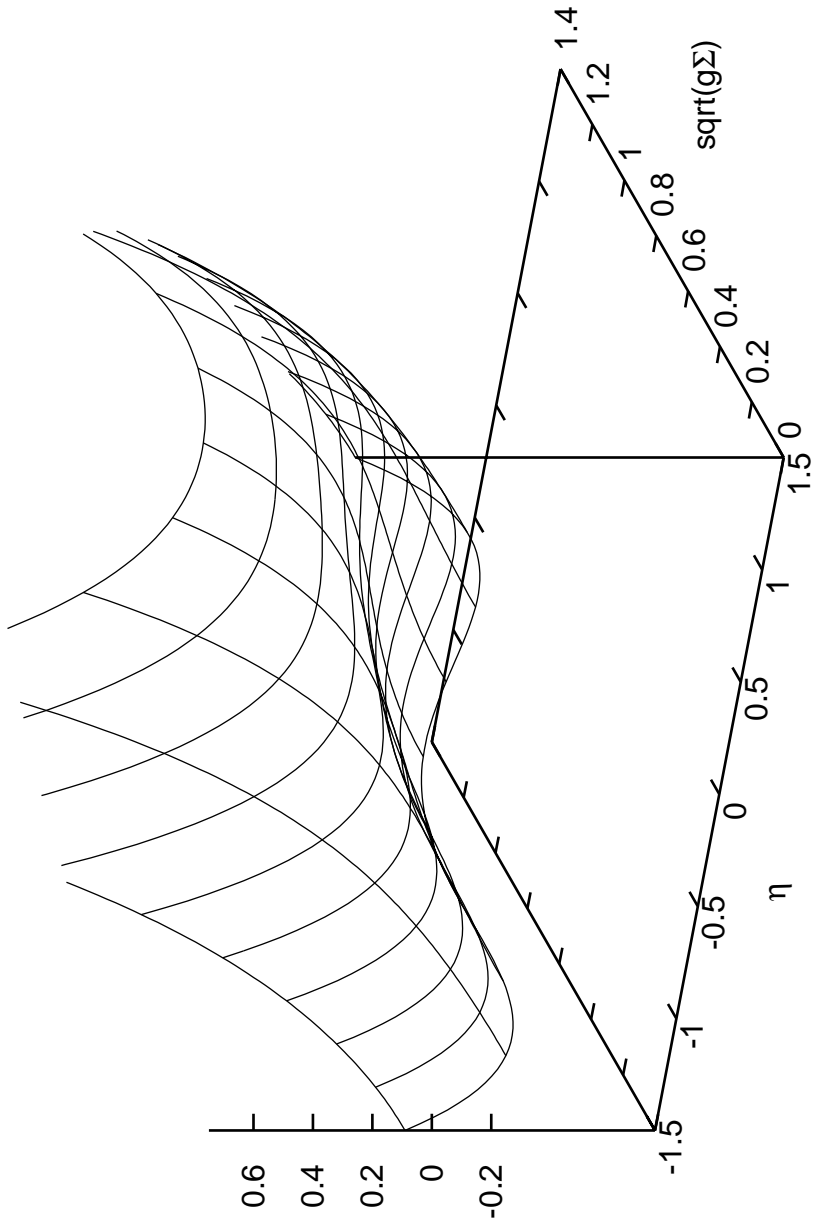}
\caption{$ j = 0.20 < j_c $. Plot of the dynamical effective potential
$ V_{eff \; dyn}(\eta,\Sigma) $ given by eq.(\ref{econs}). $ V_{eff \;
dyn}(\eta,\Sigma) $ is bounded from above by the total energy. This constrains
the trajectory represented in Fig. \ref{fetagsljc}. }
\nlabel{fsemimex}
\end{figure}

\end{document}